\documentclass{aa}
\usepackage{longtable,graphicx,amssymb,natbib,rotating,psfrag,amsmath,epsfig,array,txfonts,subfigure,multirow}

\begin{document}

\title{GRB\,070707: the first short gamma-ray burst observed by \textit{INTEGRAL}\thanks{Based on
    observations with \textit{INTEGRAL}, an ESA project with instruments and
    science data centre funded by ESA member states (especially the PI
    countries: Denmark, France, Germany, Italy,
    Switzerland, Spain), Czech Republic and Poland, and with the participation
    of Russia and the USA.}}

\author{S.~McGlynn\inst{1,2} 
\and S.~Foley\inst{1} 
\and S.~McBreen\inst{3} 
\and L.~Hanlon\inst{1} 
\and R.~O'Connor\inst{1}
\and A.~Martin Carrillo\inst{1}
\and B.~McBreen\inst{1}}

\offprints{S.~McGlynn, \email{smcglynn@particle.kth.se}}

\institute{UCD School of Physics, University College
  Dublin, Dublin 4, Ireland
   \and Department of Physics, Royal Institute of Technology (KTH), AlbaNova University Center, SE-10691 Stockholm, Sweden
\and Max-Planck-Institut f\"{u}r extraterrestrische Physik, D-85741 Garching, Germany}

\date{Received \textit{20 December 2007} / Accepted \textit{09 May 2008}}
\abstract
{\textit{INTEGRAL} has observed 47
long-duration GRBs ($T_{90}\gtrsim2$\,s) and 1 short-duration GRB
($T_{90}\lesssim2$\,s) in five years of observation since October 2002. }
{This work presents the properties of the prompt emission of GRB\,070707, which is the first short hard GRB observed by \textit{INTEGRAL}.} 
{The spectral and temporal properties of GRB\,070707 were determined using the two sensitive coded-mask $\gamma$-ray instruments on board \textit{INTEGRAL}, IBIS and SPI. } 
{The T$_{90}$ duration was 0.8\,s, and the spectrum of the prompt emission was obtained by joint deconvolution of IBIS and SPI data to yield a best fit power-law with photon index $\alpha = -1.19\,^{+0.14}_{-0.13}$, which is consistent with the characteristics of short-hard $\gamma$-ray bursts. The peak flux over 1 second was $1.79\,^{+0.06}_{-0.21}$ photons\,cm$^{-2}$\,s$^{-1}$ and the fluence over the same interval was $(2.07 ^{+0.06}_{-0.32}) \times 10^{-7}$ erg\,cm$^{-2}$ in the energy range 20--200\,keV. The spectral lag measured between 25--50~keV and 100--300~keV is 20 $\pm$ 5~ms, consistent with the small or negligible lags measured for short bursts.}
{The spectral and temporal properties of GRB\,070707 are comparable to those of the short hard bursts detected by other $\gamma$-ray satellites, including BATSE and \textit{Swift}. We estimate a lower limit on the Lorentz factor $\Gamma \gtrsim 25$ for GRB\,070707, assuming the typical redshift for short GRBs of $z=0.35$. This limit is consistent with previous estimates for short GRBs and is smaller than the lower limits of $\Gamma \gtrsim 100$ calculated for long GRBs. If GRB\,070707 is a member of the recently postulated class of short GRBs at $z \sim 1$, the lower limit on $\Gamma$ increases to $\Gamma \gtrsim 35$.} 


\titlerunning{GRB\,070707: the first short gamma-ray burst observed by  \textit{INTEGRAL}}
\authorrunning{McGlynn et al.}

\maketitle
\section{Introduction\label{intro}}
Two different types of progenitor are thought to be responsible for short and long gamma ray bursts (GRBs). Short GRBs can be produced by the merger of two compact objects (e.g. neutron star--neutron star (NS--NS) or NS--black hole (BH), \citet{lee2007}) while the core collapse of a massive star may give rise to a long duration GRB \citep{woosley}. The merger of two neutron stars produces a rapidly spinning BH with huge energy reservoirs, orbited by a neutron-rich high density torus \citep{rosswog03a}. The binding energy of the accretion disk and the spin energy of the BH represent the two main energy reservoirs. The conventional view is that the released energy is quickly and continuously transformed into a radiation-dominated fluid, with a high entropy per baryon. This fireball is then collimated into a pair of jets, similar to the long GRB model.

Host galaxies of short GRBs include both early and late type galaxies, as well as field and cluster galaxies \citep[e.g.][]{prochaska2006}. In contrast, the host galaxies of long GRBs are typically dwarf starburst galaxies with sub-solar metallicites \citep[e.g.][]{bloom2002,ceron,sav08}. Short GRB host galaxies have lower specific star formation rates and higher metallicity than long GRB hosts \citep{berger08}. Similar types of physical processes are involved in the generation of both short and long duration GRBs \citep{nakar2007b}. The relativistic flow must be dissipated for the energy to be released in the form of prompt and afterglow emission. This dissipation is generally in the form of collisionless shocks. Most models are based on synchrotron radiation from relativistic electrons accelerated within these shocks \citep{rees94}. ``Jets" of collimated emission are implied by the observations of chromatic breaks in the lightcurve and decrease the amount of energy required in the GRB from $\sim10^{53}$ ergs to $\sim10^{50}$ ergs. The central engine or progenitor of a GRB must be a compact object, e.g. a BH or NS, for such a large amount of energy ($\sim 0.1$ M$_{\odot}$) to be released over such a short timescale. Such a violent process implies the collapse of a massive star and the birth of the compact object. 

Major progress in the field of short GRBs has been made with the launch of \textit{Swift} in late 2004. \textit{Swift} has observed 26 short bursts to the end of 2007, with X-ray and/or optical afterglows detected for $\sim$\,65\% of the sample. The temporal decay slopes of the X-ray afterglows range from $\delta \sim -1$ to $\delta \sim -6$. If short GRBs are associated with NS--NS/NS--BH binary mergers, no supernova (SN) detections are expected. The recent discovery of the first X-ray afterglow of a short burst, GRB\,050509b, led to an association with an elliptical galaxy at $z=0.225$ \citep{gehrels05}. The optical afterglow of the short burst GRB\,050709 observed by HETE-II at $z=0.16$ \citep{villas2005,fox2005} ruled out a supernova association in the optical lightcurve down to a limiting magnitude of $M_R >12$, indicating that short GRBs may have a different physical origin to the majority of long GRBs. 

To complicate the picture, however, two GRBs were recently discovered with no evidence of any supernova accompanying them to deep limits. GRB\,060505 \citep{fynbo06} and GRB\,060614 \citep{galyam,dellav06} were both observed by \textit{Swift} at low $z$. The upper limits were far below the fluxes of any previous type Ic SNe \citep{fynbo06}. Both GRBs were located in star-forming galaxies \citep{watson07}, similar to long GRB host galaxies.

GRB\,060505 had a T$_{90}$ of 4\,s, while GRB\,060614 had a T$_{90}$ of 102\,s, consisting of an initial hard pulse of 5\,s with a lag of $3 \pm 6$\,ms, followed by softer emission. This led to the hypothesis that GRB\,060614 was in fact a short GRB with extended soft emission \citep{gehrels2006} or a member of an entirely new sub-class of SN-less GRBs. However, extensive spectral lag analysis has shown that the lag of GRB\,060505 is inconsistent with that of short GRBs, leading to the hypothesis that GRB\,060505 was the product of a ``failed" SN caused by the fallback of matter onto a BH \citep{sf08}.

The T$_{90}$ distribution of the BATSE GRB catalogue is bimodal \citep{kouv1993} and lognormal \citep{mcbreen1994}, with a dip at $\sim\,$2\,s and an overlap in the distribution, so that not all short bursts have a duration $<$\,2\,s. The width of each pulse in the burst lightcurve is much smaller in short GRBs than in long GRBs \citep{mcbreen03}. Long soft bumps following the initial hard pulse have been observed in the lightcurves of some \textit{Swift} short GRBs \citep[e.g.][]{fox2005,norris2006} and there is speculation that this is a characteristic feature of many short bursts.

A refinement in burst classification can now be made using a combination of $\gamma$-ray properties and environmental properties of the burst region that result from extensive observations of GRB host galaxies \citep{donaghy2006}. These include the duration of the prompt emission, spectral hardness, spectral lag, beamed equivalent radiated energy (E$_{\gamma}$), host galaxy type, location within the host galaxy and the existence of a long soft bump following the hard emission in some cases. 

The hardness ratios of short GRBs are larger on average than long GRBs \citep{kouv1993}. \cite{ghir2004} compared the spectra of short BATSE bursts with a peak flux above 10 photons cm$^{-2}$ s$^{-1}$ with the spectra of long GRBs analysed by \cite{ghir2002} and found that short bursts are, on average, spectrally harder than long bursts because of a harder photon index $\alpha$, but have similar spectral properties to the first 2~seconds of long bursts.

The spectral lag is a measure combining the temporal and spectral properties of the prompt $\gamma$-ray emission \citep[e.g.][]{norris2000,sf07}. A positive lag value indicates hard-to-soft evolution \citep{dan2003}, i.e. high energy emission arrives earlier than low energy emission. The distributions of spectral lags of short and long GRBs are noticeably different, with the lags of short GRBs concentrated in the range $\pm$ 30~ms \citep[e.g.][]{norris2006,yi2006}, while long GRBs have lags covering a wide range with a typical value of 100\,ms \citep[e.g.][]{hakkila}. Short hard bursts also tend to have lower luminosities than classical long GRBs, and therefore are not consistent with the anti-correlation observed between lag and luminosity for long GRBs \citep{norris2002}.

The beamed equivalent $\gamma$-ray energy, E$_{\gamma}$, is lower for short GRBs (10$^{48}$ -- 10$^{49}$ erg) than long GRBs (10$^{50}$ -- 10$^{51}$ erg). Short GRBs are inconsistent with the relation between the peak energy E$_{peak}$ of the $\nu$F$_{\nu}$ spectrum, and the source frame isotropic energy E$_{iso}$ \citep{amati02}, but may be consistent with the E$_{peak}$--E$_\gamma$ relation \citep{ghirlanda:2004}.

\textit{INTEGRAL} \citep{wink2003} has observed 47 long-duration GRBs ($T_{90}\gtrsim2$\,s) and 1 short-duration GRB
($T_{90}\lesssim2$\,s) in five years of observation (October 2002--December 2007) including the intense burst GRB\,041219a \citep{mcbreen06,me07} and two optically ``dark" bursts, GRB\,040223 and GRB\,040624 \citep{fill06}. The \textit{INTEGRAL} satellite consists of two coded mask $\gamma$-ray instruments, a spectrometer (SPI, \citet{ved2003}), an imager (IBIS, \citet{uber2003}) and two smaller instruments, a coded mask X-ray monitor (JEM--X, \citet{lund2003}) and an optical camera (OMC, \citet{mashesse2003}). SPI operates in the energy range 20~keV--8~MeV and IBIS in the energy range 15~keV--10~MeV. The first \textit{INTEGRAL} GRB catalogue is presented in \citet{sf07}. 

In this paper, we present the prompt and afterglow properties of the first short GRB detected by \textit{INTEGRAL}, GRB\,070707. The prompt temporal and spectral properties from SPI and IBIS are presented in Sections \ref{t} and \ref{s}. The properties of the X-ray and optical afterglow emission are presented in \S\ref{xrt} and \S\ref{opt} respectively. We discuss the implications of these results in \S\ref{disc}. 

The cosmological parameters adopted throughout the paper are $H_0$ = 70\,km\,s$^{-1}$\,Mpc$^{-1}$, $\Omega_m = 0.3$, $\Omega_{vac} = 0.7$. We adopt the notation for the $\gamma$--ray spectra that $\alpha$ represents the power--law photon index, $E_{peak}$ is the peak energy of the cutoff power-law fit and $E_0$ is the exponential rolloff energy. The power--law photon index of the X--ray spectrum is represented by $\beta$. All errors are quoted at the 1$\sigma$ confidence level.

\section{Observations\label{obs}}
 The short hard GRB\,070707 was detected by the \textit{INTEGRAL} Burst Alert System (IBAS, \citet{mere03}) at
16:08:38 UT on 7 July 2007 \citep{gcn6605} at an instrument off-axis angle of
12$^{\circ}$, within the partially coded field of view of IBIS. It was localised to R.A = 17h 51m 00.14s, Dec =  $-$68$^{\circ}$ 54' 51.8" with an uncertainty of 2.1' at the 90\% confidence level \citep{gcn6607}. The burst was also observed by KONUS--Wind \citep{gcn6615} which reported a fluence from 20~keV to 2~MeV of $(1.41 ^{+0.16}_{-1.07}) \times 10^{-6}$ erg cm$^{-2}$. The X--Ray Telescope (XRT, \citet{burrows}) on \textit{Swift} \citep{SWIFT} observed the burst location, detecting its X-ray afterglow, starting about 9 hours post-trigger \citep{gcn6610} and subsequently from $325-517$ ks after the trigger \citep{gcn6626}. Follow-up optical observations were carried out by the Ultraviolet/Optical Telescope (UVOT, \citet{roming05}) 8.8 hours after the burst \citep{gcn6611}, yielding an upper limit of $V>19.7$. A fading optical afterglow was detected with the VLT about 11 hours after the trigger \citep{gcn6609,gcn6612,gcn6613}.

\section{Properties of the prompt emission}
\subsection{Temporal Analysis\label{t}}
\begin{figure}[t]
\begin{center}
\includegraphics[width=0.6\columnwidth,angle=-90]{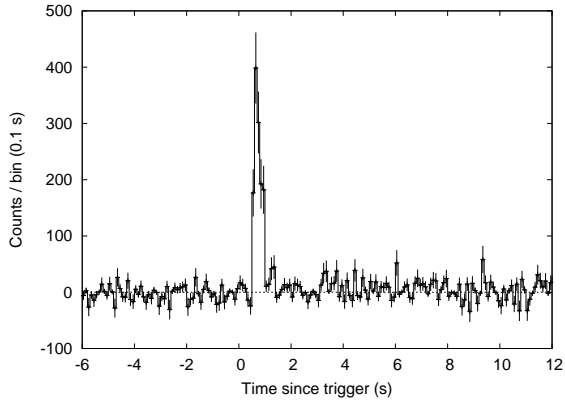}
\caption{IBIS/ISGRI lightcurve of GRB\,070707 in 0.1 second bins in the energy range 20--200\,keV. The trigger time for the burst is 16:08:38 UTC. \label{lc_0707}}
\end{center}
\end{figure}

The T$_{90}$ duration was determined to be 0.8 $\pm$ 0.2 seconds, using the 0.1 second resolution data from IBIS. Fig.~\ref{lc_0707} presents the GRB lightcurve in the energy range 20--200~keV.

The spectral lag and associated error were measured between the energy ranges 25--50~keV and 100--300~keV using the cross-correlation technique described in \cite{norris2000}. The lightcurve data were rebinned to compute the lag and errors over a higher time resolution than the natural binning of the raw data. The lag was determined to be 20 $\pm$ 5~ms from the maximum of the fourth order polynomial fit to the cross-correlation function.

 There are signs of a bump at $\sim 3-7$ seconds after the trigger in the IBIS 20--200~keV lightcurve (Fig.~\ref{lc_0707}). This is not observed in the 50--200\,keV KONUS--Wind lightcurve\footnote{http://gcn.gsfc.nasa.gov/konus{\_}grbs.html}. The IBIS lightcurve data between 20--30~keV was visually examined and no soft tail was found.

\subsection{Spectral Analysis\label{s}}
The spectrum of GRB\,070707 was extracted from the spectrometer SPI and the low energy detector ISGRI of the imager IBIS using Online Software Analysis version 5.0. The data from both instruments were fit simultaneously using XSPEC v11.3.2 and significant emission was detected up to 400\,keV. The spectra were rebinned to have a minimum of 20 counts/bin. The best fit model was determined to be a simple power--law.

\begin{table}[t]
\caption{Spectral parameters for GRB\,070707 fit by a simple power-law model for IBIS/ISGRI and SPI and a joint fit to both instruments. The columns list the values for the photon index $\alpha$, the best fit reduced $\chi^2_r$ per degrees of freedom (DoF), and fluence in the energy range 20--200~keV. The final two rows list the best fit parameters of the cutoff power-law model when fit to the joint data.
\label{spec}}
\centering
\begin{tabular}{l|ccl}
\hline\hline
\it Interval & \it $\alpha$ & \it $\chi^2_r$ &  \it Fluence  \\
& & \it /DoF & \it (20--200\,keV) \\
& & & \it ($\times 10^{-7}$ erg cm$^{-2}$) \\
\hline\hline
 PL Fit: & & & \\
Rise Time (ISGRI) & $-1.26~^{+0.21}_{-0.20}$ & 1.25/24 & $2.04~^{+0.06}_{-0.52}$\\
Fall Time (ISGRI) & $-1.36~^{+1.24}_{-1.22}$ & 1.14/6 & $0.44~^{+0.05} _{-0.36}$ \\
T$_{90}$ (ISGRI) & $-1.19~^{+0.19}_{-0.18}$ & 1.24/26 & $1.98~^{+0.05}_{-0.43}$ \\
T$_{90}$ (SPI) & $-1.19~^{+0.23}_{-0.20}$ & 0.59/15 & $2.58~^{+0.10}_{-0.75}$ \\
T$_{90}$ (Joint Fit)& $-1.19~^{+0.14}_{-0.13}$ & 1.09/47 & $2.07~^{+0.06}_{-0.32}$ \\
& & &\\
\hline
& & &\\
Cutoff PL Fit: & $-1.04~^{+0.26}_{-0.33}$ & 1.11/45 & $2.11~^{+0.11}_{-0.81}$ \\
(T$_{90}$, Joint Fit)&  \multicolumn{3}{l}{$E_0 = 673~^{+786}_{-270}$\,keV, $E_{peak} = 645~^{+766}_{-330}$\,keV} \\
& & & \\
\hline
\end{tabular}
\end{table}

The burst was divided into two parts comprising the rise-time and fall-time and separate spectra were generated. The spectral fits of both intervals were consistent within the errors, indicating that there was no significant spectral evolution in the burst. Table~\ref{spec} lists the best fit spectral parameters to the rise time, fall time and T$_{90}$ of GRB\,070707. The peak flux over the T$_{90}$ interval was $1.79\,^{+0.06}_{-0.21}$ photons\,cm$^{-2}$\,s$^{-1}$ in the energy range 20--200\,keV, and $2.19\,^{+0.05}_{-0.34}$ photons\,cm$^{-2}$\,s$^{-1}$ in the energy range 20--400~keV, i.e. the limit of the significant emission.

\begin{figure}[t]
\begin{center}
\includegraphics[width=0.6\columnwidth,angle=-90]{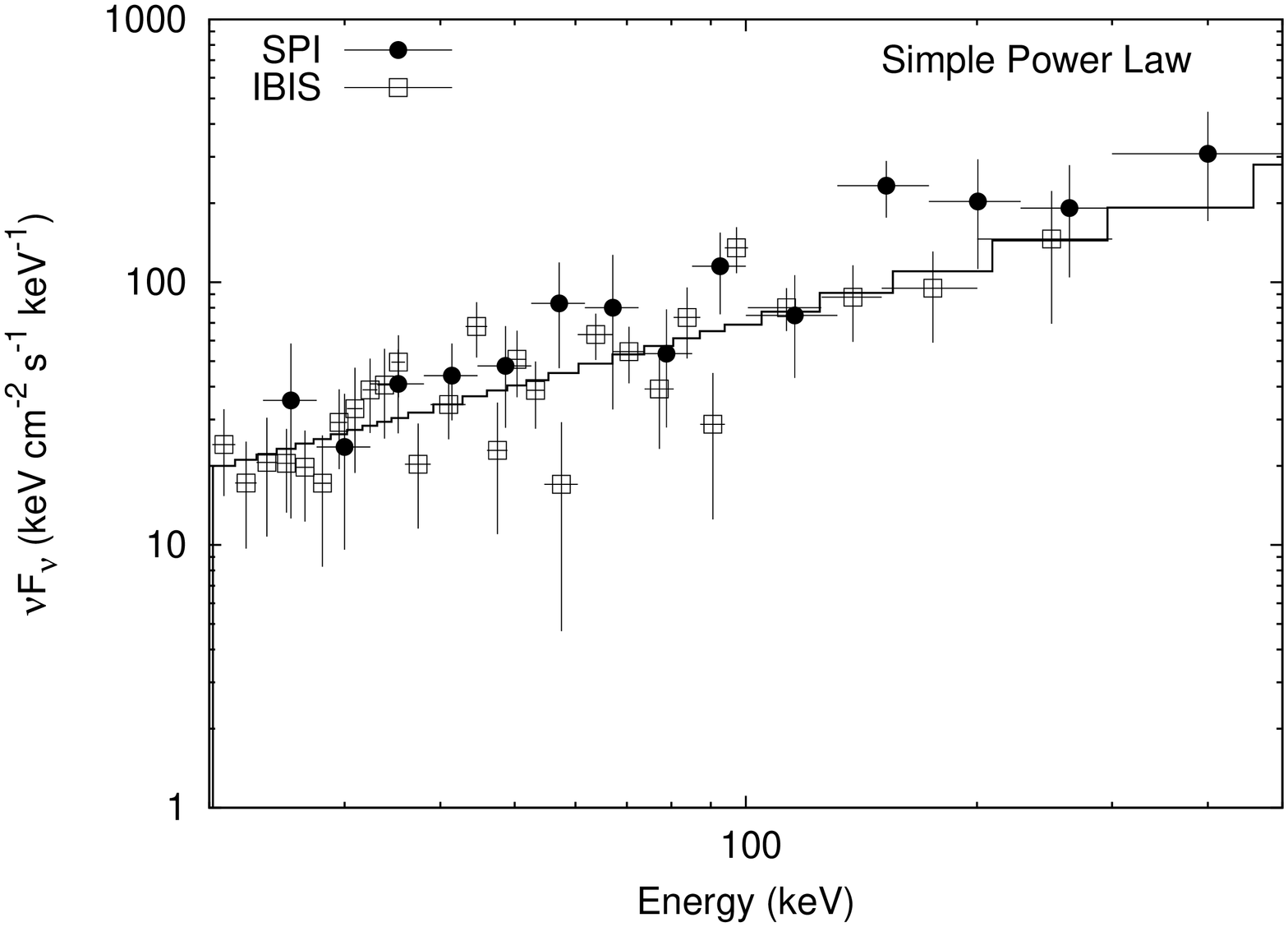}
\includegraphics[width=0.6\columnwidth,angle=-90]{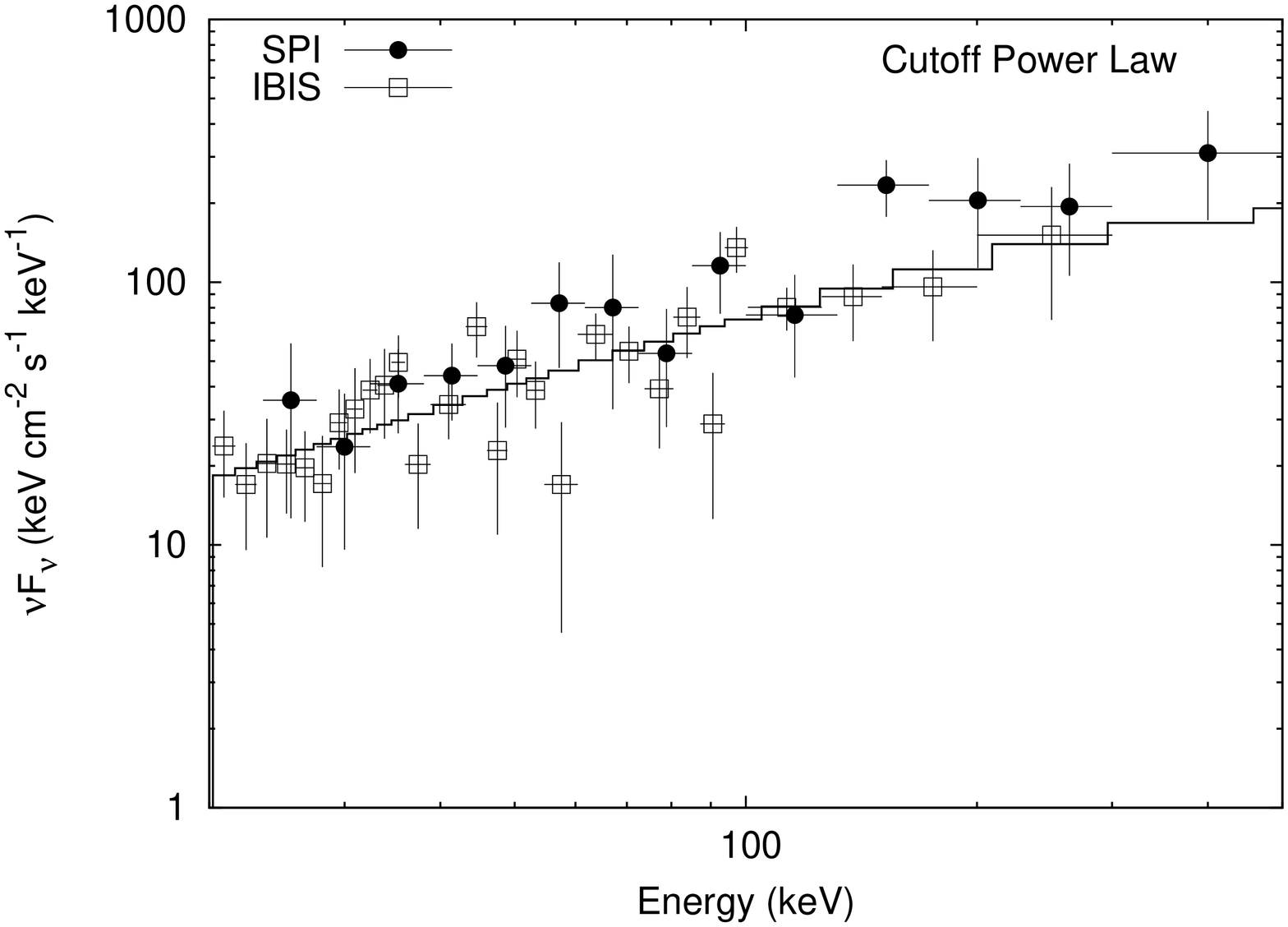}
\caption{The $\nu$F$_{\nu}$ combined IBIS/ISGRI (blue squares) and SPI (red circles) spectra of GRB\,070707 with the best fit from a simple power-law model (upper panel), with $\alpha = -1.19 ~^{+0.14}_{-0.13}$ and a cutoff power-law model (lower panel) with $\alpha = -1.04~^{+0.26}_{-0.33}$ and $E_{peak} = 645~^{+766}_{-330}$\,keV. \label{070707_spec}}
\end{center}
\end{figure}

The joint fit from IBIS/ISGRI and SPI agreed closely with the data from each instrument but with smaller error bars. The best fit model to the spectrum over the T$_{90}$ interval is a simple power-law model with photon index $\alpha = -1.19~^{+0.14}_{-0.13}$ and is shown in Fig.~\ref{070707_spec}. The fluence over the T$_{90}$ interval was $(2.07 ^{+0.06}_{-0.32})\,\times 10^{-7}$ erg cm$^{-2}$ in the energy range 20--200\,keV, and $(3.90 ^{+0.18}_{-0.71})\,\times 10^{-7}$ erg cm$^{-2}$ in the energy range 20--400\,keV. 

\cite{gcn6615} reported a cutoff power-law fit to the KONUS--Wind data with photon index $\alpha = -0.57~_{-0.43}^{+0.59}$ and peak energy E$_{peak} = 427~^{+374}_{-144}$~keV in the 20--2000\,keV energy range. The fluence obtained by KONUS--Wind in the same range was $(1.41 ^{+0.16}_{-1.07}) \times 10^{-6}$ erg cm$^{-2}$.  A cutoff power-law fit was also applied to the \textit{INTEGRAL} data for comparison (last 2 rows, Table~\ref{spec}) and the results are consistent within the errors with the KONUS--Wind parameters. Statistically this fit was poorer than the simple power-law model, which remained the best fit overall. The parameters from the cutoff power-law model are used in \S\,\ref{disc} to estimate a lower limit on the Lorentz factor of the $\gamma$-ray source. A blackbody + power-law model was fit to the data, but the contribution of the thermal component to the overall fit was almost nonexistent, indicating that GRB\,070707 has a non-thermal spectrum. The Band model \citep{band:1993} was also fit to the data, but the fit was poorly constrained.

\section{Afterglow X--ray Observations\label{xrt}}
\textit{Swift} did not trigger on GRB\,070707 since the burst occurred outside the BAT field of view and no observations were made until 8.8 hours after the burst \citep{rep_75}. The X-ray Telescope (XRT) observed the GRB location and detected a source at R.A = 17h 50m 58.49s, Dec = $-$68$^{\circ}$  55' 27" with an uncertainty of 5.4" at the 90\% confidence level, within the \textit{INTEGRAL} error region. XRT carried out a further observation from $325-517$\,ks after the 
trigger, but the source had faded to below the XRT detection limit (the 3 sigma upper limit to the observed count rate was 
0.011 counts/s, \citet{gcn6626}). Therefore it was not possible to estimate a break time from the lightcurve.

The X-ray spectrum over the interval  T$_0$ + 31.8~ks to T$_0$ + 54.1~ks was fit by an absorbed power-law with a photon index $\beta$ =  $-$2.7 $\pm$ 0.6, a fixed Galactic column density of 6 $\times 10^{20}$ cm$^{-2}$ and an average observed 0.3--10~keV flux of (2.4$^{+2.0}_{-1.4})~ \times$ 10$^{-13}$ erg cm$^{-2}$ s$^{-1}$ \citep{gcn6626}.  

\section{Afterglow Optical Observations\label{opt}}
The VLT observed the error region of GRB\,070707 approximately 11 hours after the burst occurred \citep{gcn6612} and found a single source with magnitude $R \sim 23.0$ within the XRT error circle. This source faded by the second observation at $\sim$ 34 hours post-trigger and was determined to be the optical afterglow \citep{gcn6613}. An upper limit was placed on the redshift of $z < 3.6$ (S. Piranomonte, priv. comm.) due to the absence of a Lyman alpha limit. The host galaxy has an apparent magnitude of $R = 27.3$ and is the faintest host detected so far for a short burst (e.g. the host galaxy of GRB\,050709 had $R\sim21$ \citep{hjorth2005}, the host galaxy of GRB\,050724 had $K\sim15$ \citep{berger05} and the host galaxy of GRB\,060121 had $R = 26.6$ \citep{levan2006}).
  
\section{Discussion\label{disc}}
The time history and spectral properties (e.g. T$_{90}$, spectral lag and spectral shape) of the short burst GRB\,070707 are similar to those of several short BATSE bursts \citep{yuki2006} and bursts from the KONUS--Wind short GRB catalogue \citep{konus_cat}. The ratio of $\gamma$-ray fluence to X-ray fluence ($\gamma_{30-400}$/X$_{2-30}$) is $\sim$ 8, consistent with values found for BATSE short GRBs \citep{ghir2004}. There are no direct measurements of the redshift of GRB\,070707.

\subsection{Energetics}
The isotropic peak luminosity L$_{peak,iso}$ can be calculated using the 50--300\,keV peak flux of the joint data from \S~\ref{s} (P$_{50-300} = 2 \times 10^{-7}$ erg\,cm$^{-2}$\,s$^{-1}$) and assuming that GRB\,070707 is at the average redshift for short bursts, i.e. $z = 0.35$. This yields a value for L$_{peak,iso}$ of 1.1 $\times$ 10$^{50}$ ergs s$^{-1}$. The observed bolometric fluence (1--10,000\,keV) extrapolated from the power--law fit to the joint data yields an isotropic equivalent energy E$_{iso}$ of 1.8 $\times$ 10$^{51}$ ergs. The lower limit on E$_{iso}$ is 1.2 $\times$ 10$^{50}$ ergs when the observed fluence from 20--400\,keV is used.

The spectral properties of the prompt emission can be used to estimate a lower limit on the bulk Lorentz factor $\Gamma$ of the $\gamma$-ray source \citep{nakar2007}. The prompt emission of short GRBs is predominantly non-thermal, including the spectrum of GRB\,070707. This implies that the $\gamma$-ray source is optically thin to Thomson scattering of photons on $e^{+}/e^{-}$ pairs \citep{ls2001}. If the optical depth is $\tau_T < 1$, regardless of whether internal or external shocks are involved in generating the emission, a lower limit on $\Gamma$ can be estimated using the fit to the $\gamma$-ray spectrum in the equation:

\begin{equation}
 \left(\frac{{\Gamma}m_{e}c^{2}}{E_{peak}(1+z)}\right) + (4-\alpha) \rm{ln} \Gamma +\rm{ln}\left(\frac{E_{peak}}{m_{e}c^{2}}\right) \gtrsim 30
\end{equation}

where $\alpha$ is the photon index and $E_{peak}$ is the peak energy of the cutoff power-law fit. We obtain $\Gamma \gtrsim 25$ using the cutoff power-law model parameters from Table~\ref{spec} and assuming a redshift of $z=0.35$. We note that there are large errors associated with the value of $E_{peak}$, which is not well constrained by the fit to the joint data, leading to a range of $\Gamma \gtrsim 15-40$. A similar value for the limit of $\Gamma \gtrsim 20$ was obtained using the KONUS parameters, which are better constrained. This is comparable to the lower limits calculated by \citet{nakar2007} for two other short GRBs, $\Gamma \gtrsim 4$ for GRB\,050709 \citep{villas2005} and $\Gamma \gtrsim 25$ for GRB\,051221a \citep{051221a}. These model independent lower limits imply that short GRBs are ultra-relativistic, similar to long GRBs. The relatively low Lorentz factor implies a late deceleration time and a smaller initial radius, resulting in the possible late onset of the afterglow. 

The lower limits on $\Gamma$ for short GRBs are smaller than the average obtained for long GRBs of $\Gamma \gtrsim 100$ \citep{ls2001}. The bulk Lorentz factor has been estimated to be $\Gamma \sim 400$  from the early afterglows of two long bursts, GRB\,060218 and GRB\,060607a \citep{molinari07}. The Lorentz factor has also been calculated from the thermal components of the prompt emission from GRB\,970828 and GRB\,990510 at $z=0.96$ and $z=1.62$ respectively \citep{asaf07}. The calculated values ($\Gamma = 305 \pm 28, \Gamma = 384 \pm 71$, dependent on the ratio between the total fireball energy and the $\gamma$-ray energy) are consistent with the measurements of \citet{molinari07}. In general, there are fewer assumptions for the prompt emission than for the afterglow emission (e.g. microphysics parameters dependent on the environment). However, \citet{norris2006} suggested that the short spectral lags observed in short GRBs may be due to a very high Lorentz factor of $\Gamma \sim 500-1000$, since a large $\Gamma$ is necessary to avoid a significant contribution to the lag from the pulse duration due to relativistic beaming.

\citet{berger2007} recently reported host galaxy observations for 9 short GRBs. Eight of the nine are faint ($R \sim 23$), indicating the possible existence of a population of short GRBs at $z \sim 1$. This was confirmed by \citet{cenko08}, who identified two \textit{Swift} short bursts with host galaxies at $z \sim 0.9$, GRB\,070429B and GRB\,070714B. This implies that the energy release of short GRBs may be higher than previously thought and in the same range as long GRBs. The isotropic energy calculated for GRB\,070707 is 1.8 $\times$ 10$^{51}$ ergs at $z=0.35$, and 1.5 $\times$ 10$^{52}$ ergs at $z =1$, extrapolated to the 1--10,000\,keV energy range, closer to the values expected from long duration bursts \citep{butler07}.

\citet{berger2007} estimated that the expected median redshift of host galaxies with $25 < R <27$ from the Hubble Deep Field Survey \citep{coe2006} is about 1.1, so it is possible that GRB\,070707 has a redshift close to 1. If this is the case, the lower limit on the Lorentz factor at $z=1$ increases to $\Gamma \gtrsim 35$, remaining comparable to the previous limits obtained for short GRBs.

\subsection{Compact Objects}
Short GRBs are thought to be the product of the coalescence of compact binaries, e.g. NS--NS/NS--BH mergers. The duration of the GRB is determined by the lifetime of the accretion disk. Theoretical estimates yield binary merger rates that can easily accommodate the observed burst rate, with engine lifetimes and energy release roughly consistent with the burst properties for a cosmological population \citep{guetta05}. 

\citet{rosswog07} modelled the time scales and luminosities resulting from fallback in the aftermath of compact binary mergers and found that in a NS--BH merger where the masses are 1.4 and 4\,M$_{\sun}$ respectively, comparatively little fallback material is produced due to the short accretion timescale. This could produce a short GRB with low X-ray activity. A double NS merger can produce X-rays which fall off with time after an initial short-lived plateau, predicting an X-ray luminosity of L$_X \sim \frac{\epsilon_x}{0.1} \times 10^{44}$ erg\,s$^{-1}$ 1 hour after coalescence. GRB\,070707 was not observed in X-rays until $\sim$ 9 hours after the prompt emission. Assuming a redshift $z=0.35$ and using the 0.3--10~keV flux of 2.4 $\times$ 10$^{-13}$ erg cm$^{-2}$ s$^{-1}$ determined from XRT (\S\,\ref{xrt}), L$_X \sim 10^{44}$ erg\,s$^{-1}$, and might be compatible with this model.

\citet{tanvir05} reported a correlation between the locations of short bursts observed by BATSE and the positions of galaxies in the local universe. When the galaxy types are restricted to earlier morphological types, between 10\% and 25\% of short GRBs could potentially originate at low \emph{z} ($z < 0.025$). If both cosmological and local short GRBs arise from NS--NS coalescence, their association with host galaxies of intermediate age/old stellar populations is to be expected. The rate implied by \citet{tanvir05} of several bursts per year within 100\,Mpc agrees with estimates of the coalescence rate for double NS systems in our galaxy \citep{ns}.

The advanced version of LIGO could detect gravitational waves from binary systems out to $\sim$ 500\,Mpc \citep{dalal2006}. Detection of gravitational waves could provide an insight into the production mechanisms of short GRBs.

\section{Conclusions}
GRB\,070707 was the first short burst observed by \textit{INTEGRAL}, with significant emission detected up to 400\,keV. The spectral and temporal properties of GRB\,070707 are in agreement with the properties of the short hard bursts detected by other $\gamma$-ray satellites, including BATSE and \textit{Swift}. 

Assuming GRB\,070707 is at the average redshift of $z = 0.35$ obtained for short GRBs, the isotropic equivalent energy E$_{iso}$ (1--10,000\,keV) was estimated to be 1.8 $\times$ 10$^{51}$ ergs, or 1.2 $\times$ 10$^{50}$ ergs when the observed fluence from 20--400\,keV is used. The peak luminosity $L_{peak,iso}$ (50--300\,keV) was estimated to be 1.1 $\times$ 10$^{50}$ ergs s$^{-1}$. We also estimate a lower limit of $\Gamma \gtrsim 25$, consistent with previous estimates for other short GRBs and smaller than the values calculated for long duration GRBs, even when taking large errorbars into account. The relatively low Lorentz factor implies a late deceleration time and a smaller initial radius, resulting in the possible late onset of the afterglow. 

GRB\,070707 may however be a member of the recently postulated class of short GRBs at $z \sim 1$ \citep{berger2007}, since its host galaxy is faint ($R = 27.3$). At $z=1$ the isotropic equivalent energy E$_{iso}$ of GRB\,070707 is 1.5 $\times$ 10$^{52}$ ergs and the lower limit on the bulk Lorentz factor increases to $\Gamma \gtrsim 35$.

\begin{acknowledgements}
S.M.B. acknowledges
the support of the European Union through a Marie Curie Intra--European
Fellowship within the Sixth Framework Program. 
\end{acknowledgements}

\bibliography{refs}
\bibliographystyle{aa}

\end{document}